\documentclass[twocolumn,10pt]{tsfp}
\usepackage{flushend}
\usepackage{graphicx}
\usepackage[authoryear,round]{natbib}
\usepackage{fancyhdr}
 
\usepackage{amsmath}
\usepackage{paralist}
\usepackage{wrapfig}
\usepackage{enumitem}
\usepackage[export]{adjustbox}

\def\g1{0.25}
\def\gf1{0.23}

\title{THE UNIVERSALITY OF THE LAW OF THE WALL: A LONG-LASTING CONTROVERSIAL DEBATE}

\author{Stefan Heinz
    \affiliation{
	Department of Mathematics and Statistics\\
	University of Wyoming\\
	1000 E. University Avenue, Laramie, WY 82071, USA\\
    heinz@uwyo.edu
    }	
}

\begin{document}

\maketitle   
\thispagestyle{fancy}

\fontsize{9}{11}\selectfont

\section*{ABSTRACT}

The discovery of the law of the wall, the log-law including the von K\'{a}rm\'{a}n constant, is seen to be one of the biggest accomplishments of fluid mechanics. However, after more than ninety years there is still a controversial debate about the validity and universality of the law of the wall. Clarity about this question matters: in absence of alternatives, a reliable and universal theory involving the law of the wall is needed to provide essential guideline for the validation of theory, computational methods, and experimental studies of very high Reynolds number ($Re$) flows. 
The paper presents an analysis of concepts used to derive controversial conclusions. 
It is shown that nonuniversality is a consequence of simplified modeling concepts, which leads to unrealizable models. On the other hand, realizability implies universality: models in consistency with physical requirements do not need to be adjusted to different flows.
There are essential advantages of a universal law of the wall: it enables the design of accurate turbulence models and it provides a bridge between finite $Re$ observations and asymptotic structural theories of turbulence.

\section*{INTRODUCTION}

The reliable and efficient prediction of turbulent flows at high Reynolds number ($Re$), in particular the prediction of wall-bounded turbulent flows seen in reality, represents a fundamental challenge 
of computational fluid dynamics (CFD). As it is well known, direct numerical simulation (DNS), large eddy simulation (LES), and experimental studies are hardly applicable to extreme $Re$ regimes, and relatively cost-efficient hybrid RANS-LES, which combine LES with Reynolds-averaged Navier-Stokes (RANS) equations, suffer from reliability issues~\citep{JPAS-20}. There are very promising new developments as given by minimal error hybrid-RANS-LES (which can act as resolving LES)~\citep{Fluids24-CES, ApplSci24-CES, PF22, FTC-22,  PFCES-21, JPAS-20, PF20} but these methods also need evidence for their validity at high $Re$. A closely related challenge is the understanding of the nature of turbulent flow 
in the limit of infinite $Re$. 

The law of the wall, the log-law including the von K\'{a}rm\'{a}n constant~\citep{Karman-30}, is of essential relevance in this regard, it has very essential practical benefits. 
First, it can provide strict guideline for the validation of experiments and computational simulation methods such as DNS, LES, and hybrid RANS-LES for at least several canonical  high $Re$ flows~\citep{JOT-18a, JOT-18b, Fluids22-Mem}. Such validation requirement also includes recently developed highly promising minimal error simulation methods~\citep{PF22, FTC-22, PFCES-21, JPAS-20, PF20}. These methods are independent of resolution requirements of existing methods, but their suitability (using very coarse grids) needs evidence~\citep{Fluids24-CES, ApplSci24-CES}. 
Second, as is well known, the structure of usually applied turbulence models faces questions. A universal law of the wall can be applied for deriving exact turbulence models, as shown by the derivation of an exact transport equation for the turbulent viscosity~\citep{Plaut-22}.
Third, the existence of a universal law of the wall can provide essential contributions to obtaining a more comprehensive understanding of the structure of turbulent flows at high $Re$~\citep{Fluids24-As}.

But for almost a century there is still a debate about the validity and universality of the law of the 
wall~\citep{Fluids24-Law}. The specific question is whether the mean streamwise velocity $U^{+}$ of at least several canonical wall-bounded flows is characterized by log-law variations in absence of boundary effects, this means whether we have $U^{+}=\kappa^{-1} ln \:y^{+} + B$ including the same von K\'{a}rm\'{a}n constant $\kappa$ and constant $B$. 
The superscript $+$ refers to inner scaling, we use $U^{+}=U/u_{\tau}$ and 
$y^{+}=Re_{\tau}y$ for the inner scaling wall distance, where $y$ is normalized by $\delta$ which is the half-channel height, pipe radius, or 99\% boundary layer thickness with respect to channel flow, pipe flow, and the zero-pressure gradient turbulent boundary layer (TBL), respectively (the zero-pressure gradient TBL will be referred to simply as TBL). 
The friction Reynolds number is defined by $Re_{\tau}=u_{\tau} \delta / \nu$, where $u_{\tau}$ is the friction velocity  and $\nu$ is the constant kinematic viscosity.

The development of views over 80 years 
were reviewed, for example, by~\citet{Ramis-10}, \citet{Marusic-Rev}, 
\citet{Smits-Rev}, and \citet{Jimen-13}.
 Supported by increasing access to high $Re$ data, the validity and universality of the law of the wall are intensively debated over the last 15 years: traditional views in favor of universality are challenged by opposite views~\citep{Baumert-13, Kazakov-16, Hult-4,  Monke-17, Piro-23, Sreni-23,  Hansen-24, Spalart-21, Piro-23B, Cheng-23}. 
Recent analyses in favor of universality were presented, e.g., in Refs.~\citep{Luchini-17, Luchini-18, Ali-20}. 
In particular, a comprehensive analysis of available numerical and experimental data up to very high $Re$ was presented recently based on observational physics criteria~\citep{JOT-18a, JOT-18b}. 
The latter results were recently questioned by data analyses of the majority of existing DNS and experimental results, arguing in favor of nonuniversality of the law of the wall~\citep{Cant-19, Cant-22,  Monke-17, Monke-21, Monke-23}.

One motivation for this paper is to contribute to clarifying these questions by an 
identification of reasons leading to opposite conclusions regarding the universality of the law of the wall and von K\'{a}rm\'{a}n constant~\citep{Fluids24-Law}. This will be done by comparing recent analysis results in favor of nonuniversality of the law of the wall~\citep{Cant-19, Cant-22, Monke-17, Monke-21, Monke-23} with implications of observational physics criteria~\citep{JOT-18a, JOT-18b}. The concept of realizability will be involved in these discussions. In this regard, emphasis will be placed on whether or not the methods considered satisfy Reynolds stress-realizability constraints~\citep{JPAS-20} and realizability constraints arising from the entropy concept. The latter implies the need that the entropy of physically equivalent flows needs to be the same~\citep{JOT-18a, JOT-18b}. 
Another motivation for this paper is to report essential advantages of a universal law of the wall in respect to the development of turbulence models~\citep{Plaut-22} and the analysis of the asymptotic structure of wall-bounded turbulent flows~\citep{Fluids24-As}. 

\begin{figure*}[t] 
\center
\includegraphics[width=\g1\textwidth]{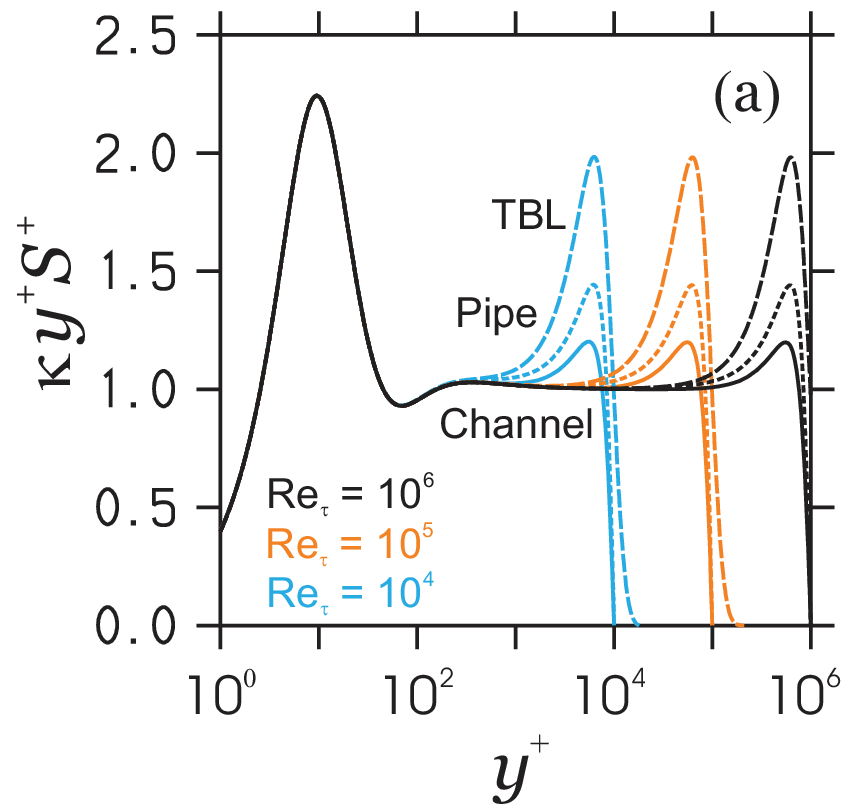} 
\includegraphics[width=\g1\textwidth]{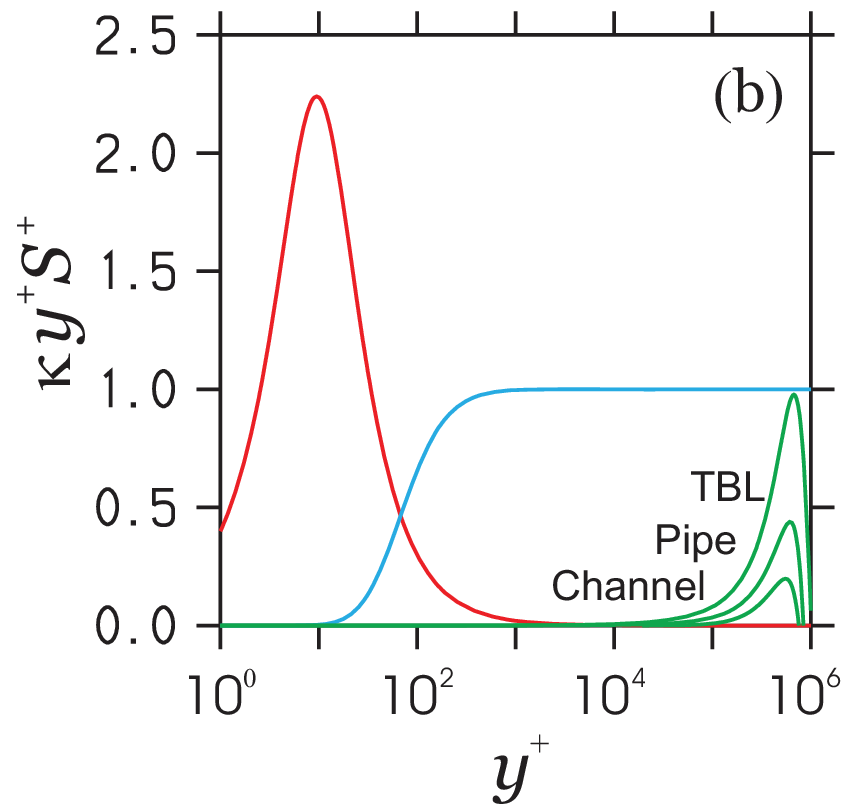} 
\includegraphics[width=\g1\textwidth]{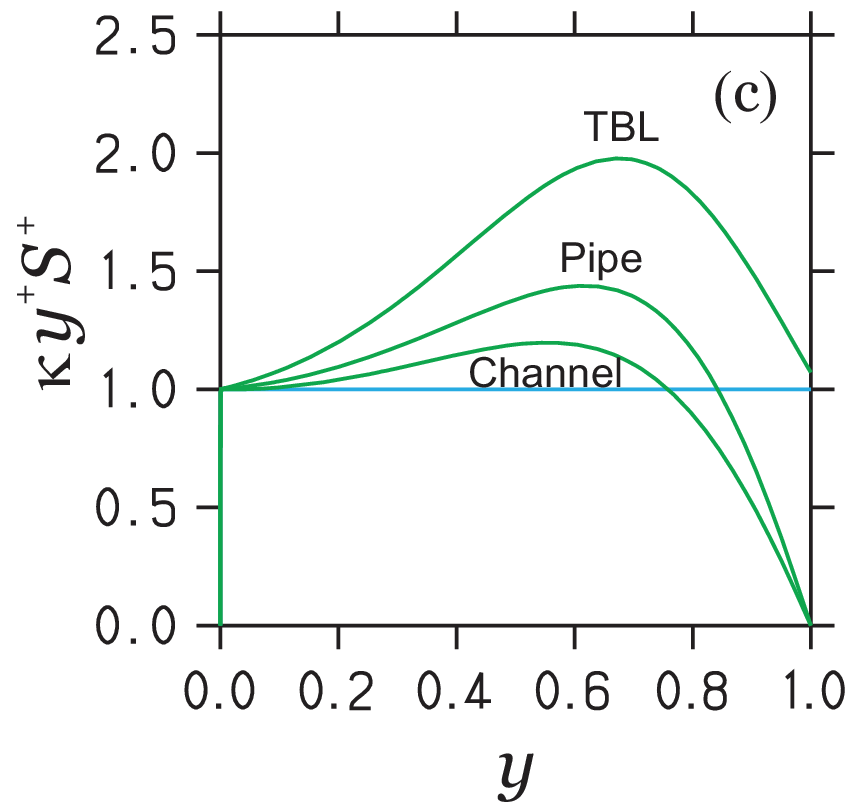} 
\linespread{1.}    \vspace{-0.3cm}
\caption{
The log-law indicator $\kappa y^{+} S^{+}$ (with $\kappa=0.4$) obtained from the PVM is shown in (a) for the given $Re_{\tau}$ and the three flows considered (channel flow: solid line; pipe flow: short dashes; TBL: long dashes). 
In (b), the mode contributions $\kappa y^{+} S_1^{+}$ (red line), $\kappa y^{+} S_2^{+}$ (cyan line), and $\kappa y^{+} S_3^{+}$ (green lines) are shown for $Re_{\tau}=10^{6}$ in inner scaling. In (c), mode contributions $\kappa y^{+} S_2^{+}$ (cyan line) and $\kappa y^{+} (S_2^{+}+S_3^{+})$ (green lines) are shown for $Re_{\tau}=10^{6}$ in outer scaling. There is no visible $\kappa y^{+} S_1^{+}$ mode.
}                                                                               
\label{fig:structure}
\vspace{0.cm}
\end{figure*}

\section*{UNIVERSAL VELOCITY MODELS}

A universal velocity model, the probabilistic velocity model (PVM) was introduced in Refs.~\citep{JOT-18a, JOT-18b} for $Re_{\tau}\geq 500$ for turbulent channel flow, pipe flow, and the TBL. It reads 
\vspace{-0.6cm}
\begin{equation} 
U^{+}=U_{1}^{+}+\frac{1}{\kappa} \:ln\left( \frac{ 1+H y^{+}/y_{\kappa} }{w+Ky} \right). 
\vspace{-0.5cm}
\label{eq:A1} 
\end{equation}
The first contribution is characterized by 
\vspace{-0.4cm}
\begin{equation}
U_{1}^{+} =\int_0^{y^{+}}\!\!\Big[1-f(t)\Big] dt, \quad  f(t)=\left[ \frac{(t/a)^{b/c}}{1+(t/a)^{b/c}} \right]^{c}, 
\label{eq:A1p} 
\vspace{-0.4cm}
\end{equation}
where $a=9, b=3.04, c=1.4$. It is worth noting that $U_{1}^{+}$ represents an analytically given function~\citep{Fluids24-Law}. The second contribution includes the universal von K\'{a}rm\'{a}n constant 
$\kappa=0.40$, $y_{\kappa}=75.8$, and $K=(0.933, 0.687, 0.285)$, where $(\cdots, \cdots, \cdots)$ refers to channel flow, pipe flow, and TBL. The functions $w=(w_{CP}, w_{CP}, w_{BL})$ and $H$ are given by 
\vspace{-0.4cm}
\begin{align*}
w_{CP}=0.1(1-y)^{2}[ 6y^{2}+11y+10],~  w_{BL}=e^{-y(0.9+y+1.09y^{2})},
\label{eq:A2p} 
\end{align*}
\vspace{-1.4cm}
\begin{equation}
H= \left[ \frac{ y^{+}/12.36   } {1+y^{+}/12.36 } \right]^{6.47}.  
\label{eq:A2} 
\vspace{-0.2cm}
\end{equation}
A closer look at the structure of the PVM shows the following. 
In addition to depending on the constants $\kappa, K, y_{\kappa}$, the PVM only depends on $f, H, 1-w$ (this explanation is simplified by taking reference to $1-w$ instead of $w$). The latter are 
are non-decreasing functions varying between zero and unity. Thus, these functions play the role of distribution functions [i.e. integrals over probability density functions (PDFs)], which characterize regime transitions. The same applies to $U_{1}^{+}/U_{1\infty}^{+}$ (where $U_{1\infty}^{+}=15.85$) implied by $f$. This means, 
 in addition to $\kappa, K, y_{\kappa}$, the PVM only depends on the regime transition control parameters $H, 1-w$, and $U_{1}^{+}/U_{1\infty}^{+}$. 

Used in conjunction with models for the total stress $M$ and the momentum balance 
$S^{+}-\langle u'v'\rangle^{+}=M$ where $S^{+}=\partial U^{+}/\partial y^{+}$, we note that the PVM also implies analytical models for the Reynolds shear stress $\langle u'v'\rangle^{+}$, turbulence production, turbulent viscosity, bulk velocity, skin-friction coefficient and bulk Reynolds number~\citep{JOT-18a, JOT-18b}. Asymptotic limits of these variables are reported elsewhere~\citep{JOT-18b}.

\begin{figure*}[t] 
\center
\includegraphics[width=\g1\textwidth]{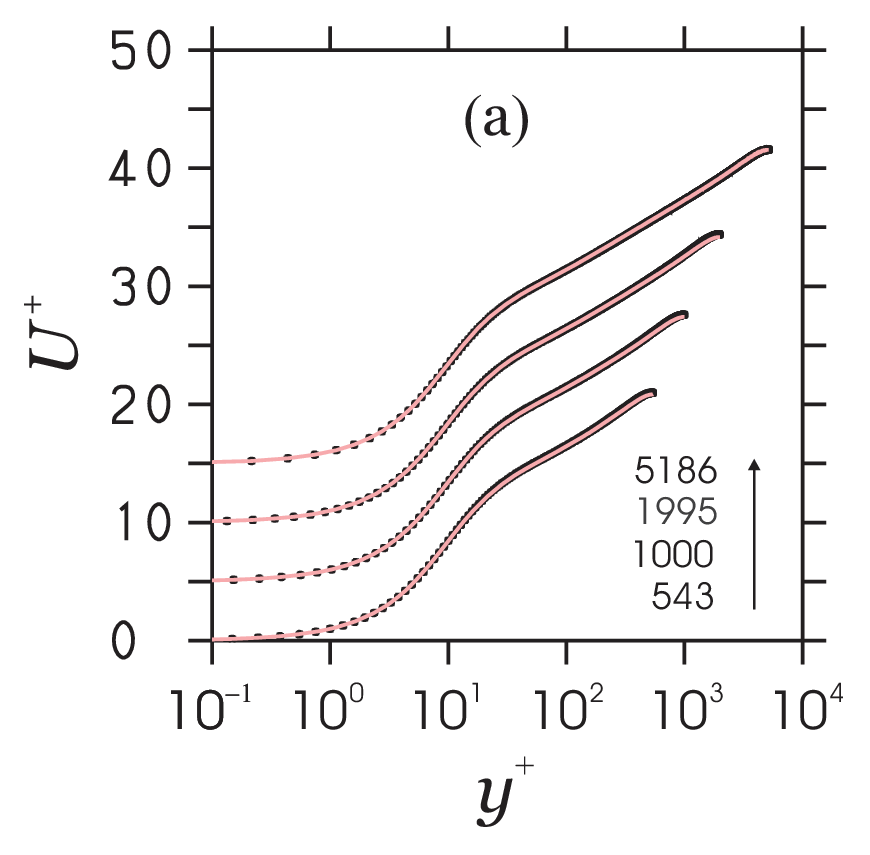} 
\includegraphics[width=\g1\textwidth]{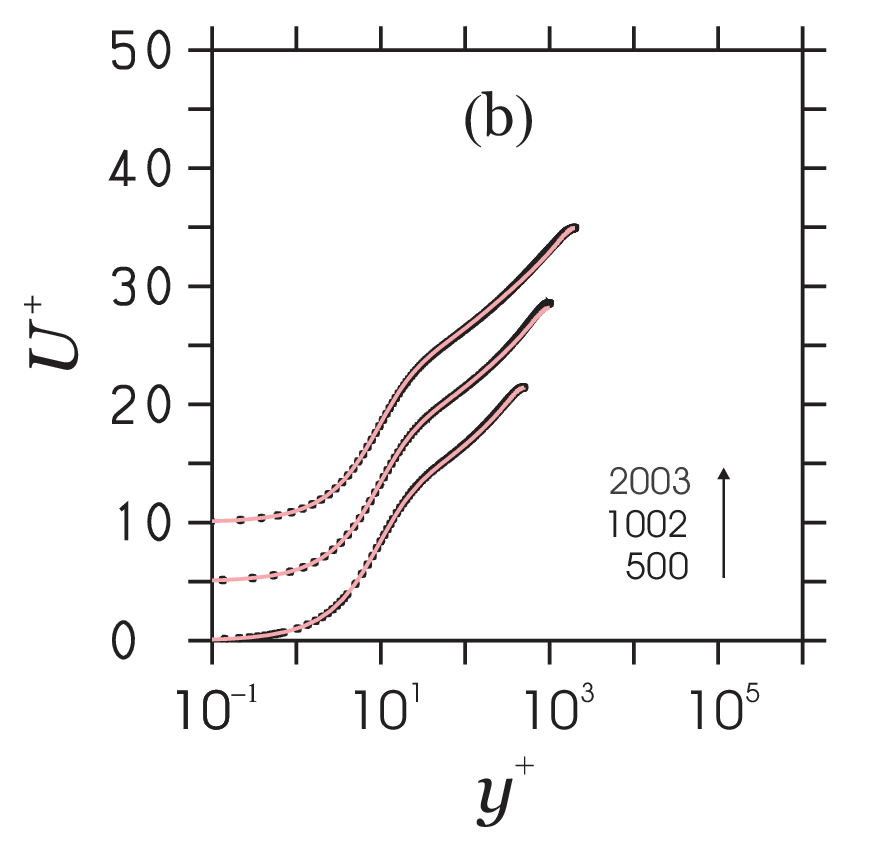} 
\includegraphics[width=\g1\textwidth]{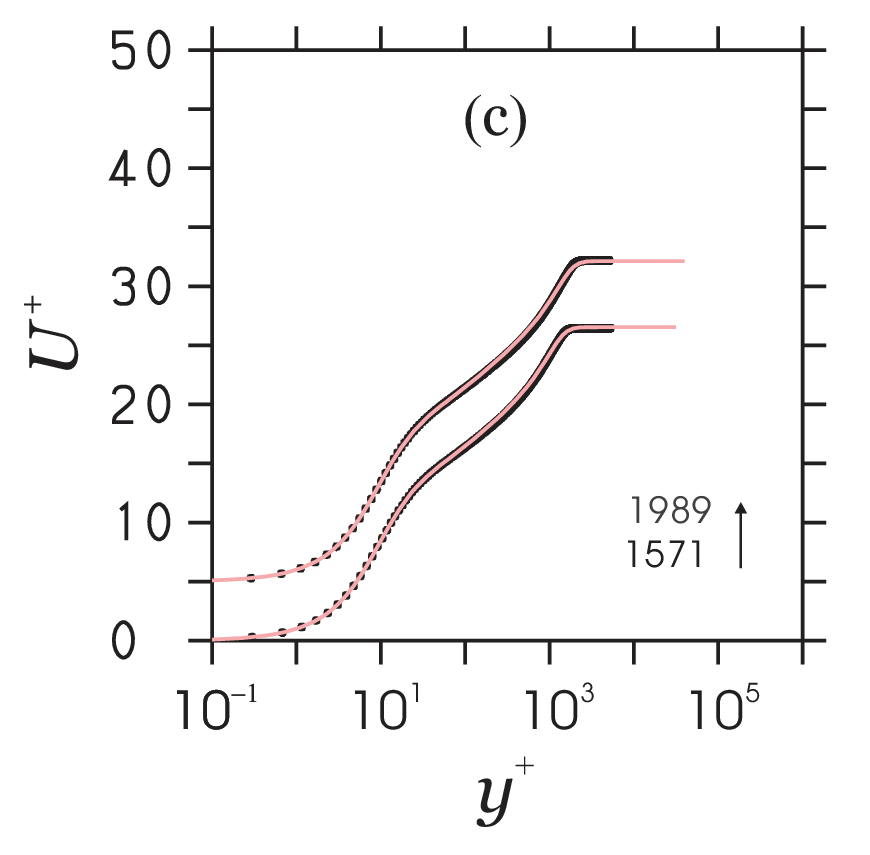} 
\linespread{1.}    \vspace{-0.3cm}
\caption{
The PVM (lines) compared to DNS data (dots) for the given $Re_{\tau}$ (separated by $\Delta U^{+}=5$). (a) Channel flow; DNS data of Lee \& Moser \citep{DNS-M}, (b) pipe flow; DNS data of Chin et al. \citep{DNS-pipe}, (c) TBL; DNS data of Sillero et al. \citep{Sillero-13}. 
}                                                                               
\label{fig:DNS}
\vspace{0.cm}
\end{figure*}

\begin{figure*}[t!] 
\center
\includegraphics[width=\g1\textwidth]{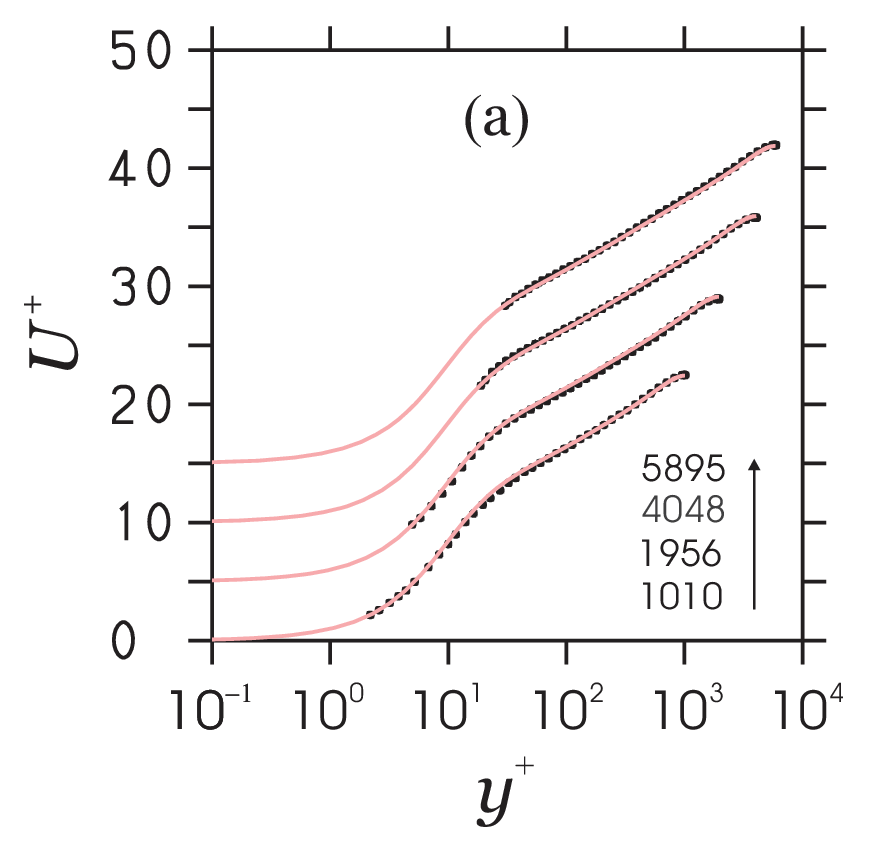} 
\includegraphics[width=\g1\textwidth]{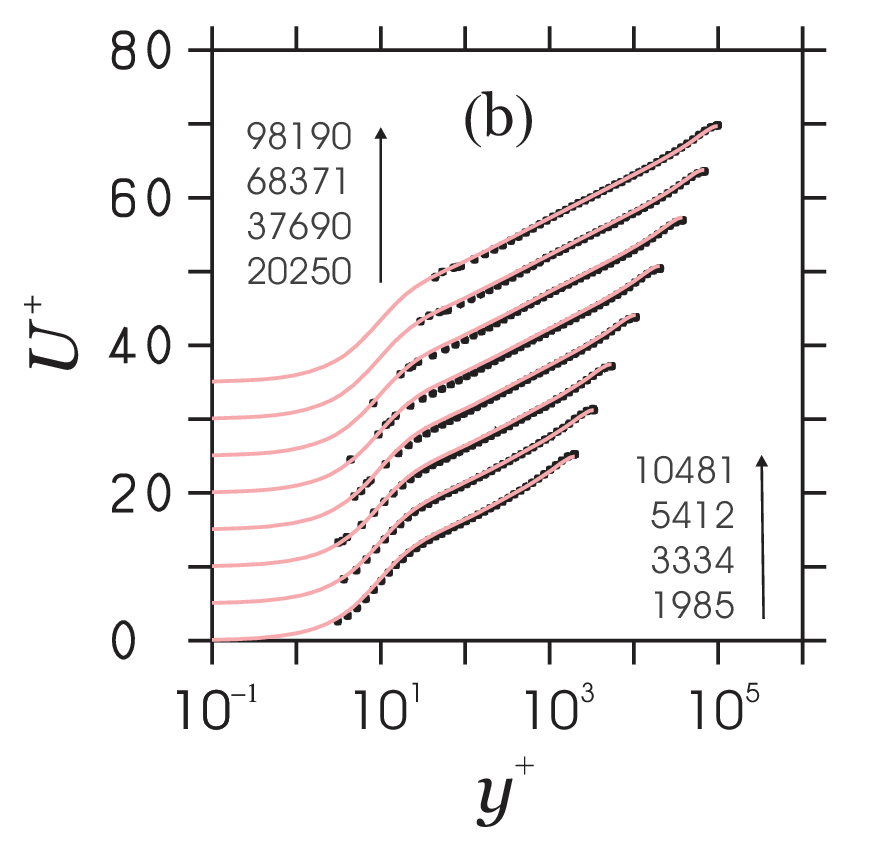} 
\includegraphics[width=\g1\textwidth]{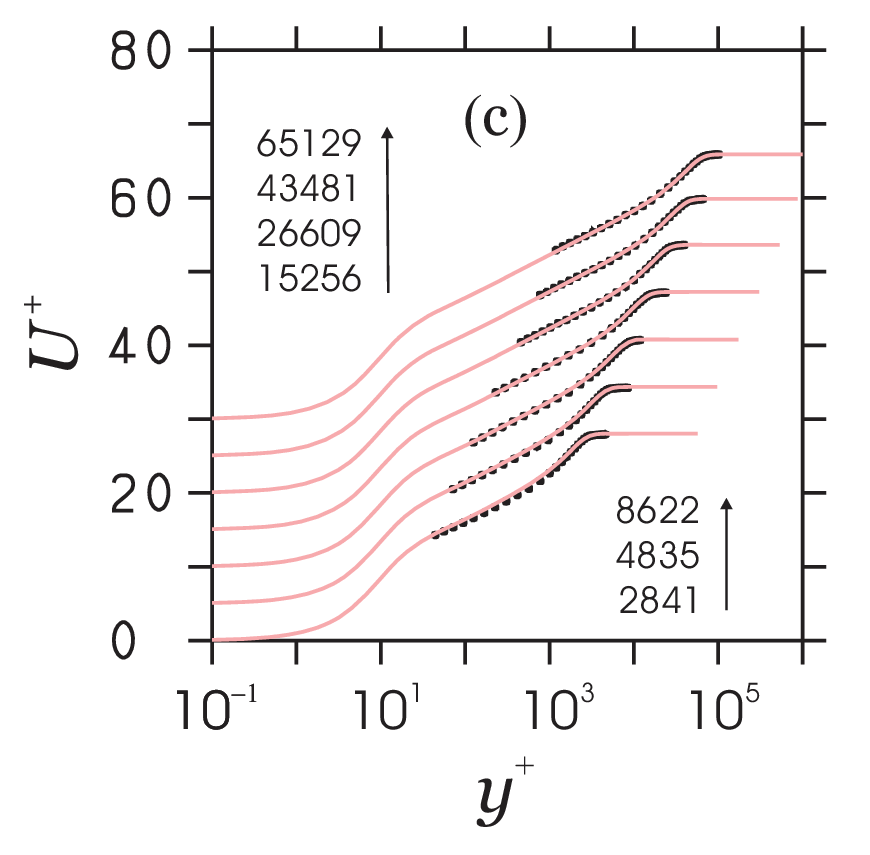} 
\linespread{1.}    \vspace{-0.3cm}
\caption{
The PVM (lines) compared to experimental data (dots) for the given $Re_{\tau}$ (separated by $\Delta U^{+}=5$). (a) Channel flow; experimental data of Schultz \& Flack \citep{Flack}, (b) pipe flow; experimental data of Hultmark et al. \citep{Hult-3}, (c) TBL; Pitot experimental data of Vallikivi et al. \citep{Valli-15}.
}                                                                               
\label{fig:Exp}
\vspace{0.cm}
\end{figure*}

Figure~\ref{fig:structure} explains the PVM structure. The model involves the contributions $S_1^{+}$, $S_2^{+}$, 
$S_3^{+}$ to the characteristic shear rate $S^{+}=\partial U^{+}/\partial y^{+}$. These shear rate contributions imply corresponding velocity contributions $U_1^{+}$, $U_2^{+}$, and $U_3^{+}$. 
The inner scaling contributions $S_1^{+}$ and $S_2^{+}$ (which are only functions of $y^{+}$) are the same for all three flows considered. The outer scaling contribution $\kappa y^{+}S_3^{+}$ (which is only a function of $y$) depends on the flow geometry. There are two inner scale correction terms $S_{1}^{CP}$ and $S_{2}^{CP}$ which ensure the correct shear rate limit at the centerline for channel and pipe flow~\citep{JOT-18a, JOT-18b}. These contributions provide insignificant corresponding contributions to the mean velocity. 
As may be seen in Fig.~\ref{fig:structure}a, the PVM clearly supports the validity and universality of the log-law. 
In absence of boundary effects, the PVM implies $ U^{+}=\kappa^{-1} ln \:y^{+}+5.03$ for all the three flows considered, where $\kappa=0.40$. A relevant conclusion of the PVM is that 
critical Reynolds numbers for the observation of a strict log-law for channel flow, pipe flow and the TBL are given by about $Re_{\tau}=20,000$, $Re_{\tau}=63,000$ and $Re_{\tau}=80,000$, respectively. 
The excellent PVM performance in comparison to DNS and high-$Re$ experimental data is illustrated in Figs.~\ref{fig:DNS}, \ref{fig:Exp}.

There are other models that support the validity and universality of the log-law. One such model is the model of 
\citet{Luchini-17, Luchini-18}. The characteristic shear rate is described by 
$S^{+}=1/(\kappa y^{+})+A_1g/Re_{\tau}$, where $\kappa=0.392$, $A_1=1$, and $g=(1, 2, 0)$ for channel flow, pipe flow, and the TBL, respectively. Luchini's model may be seen as basis of the model presented recently by Monkewitz and Nagib (MN)~\citep{Monke-23}, see the discussion below in regard to Fig.~\ref{fig:Monk3}. The difference is that Luchini does not make an attempt to introduce different von K\'{a}rm\'{a}n constants and other constants for different flows considered.

\section*{NONUNIVERSAL VELOCITY MODELS}

\begin{figure*}[t] 
\center
\includegraphics[width=\g1\textwidth]{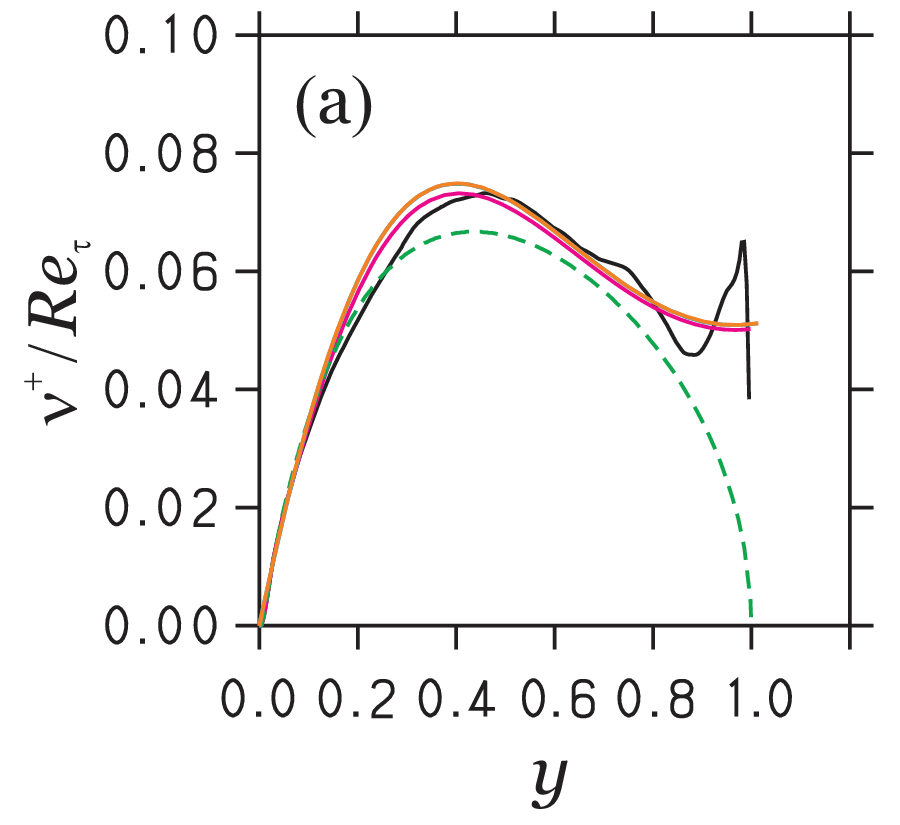} 
\includegraphics[width=\g1\textwidth]{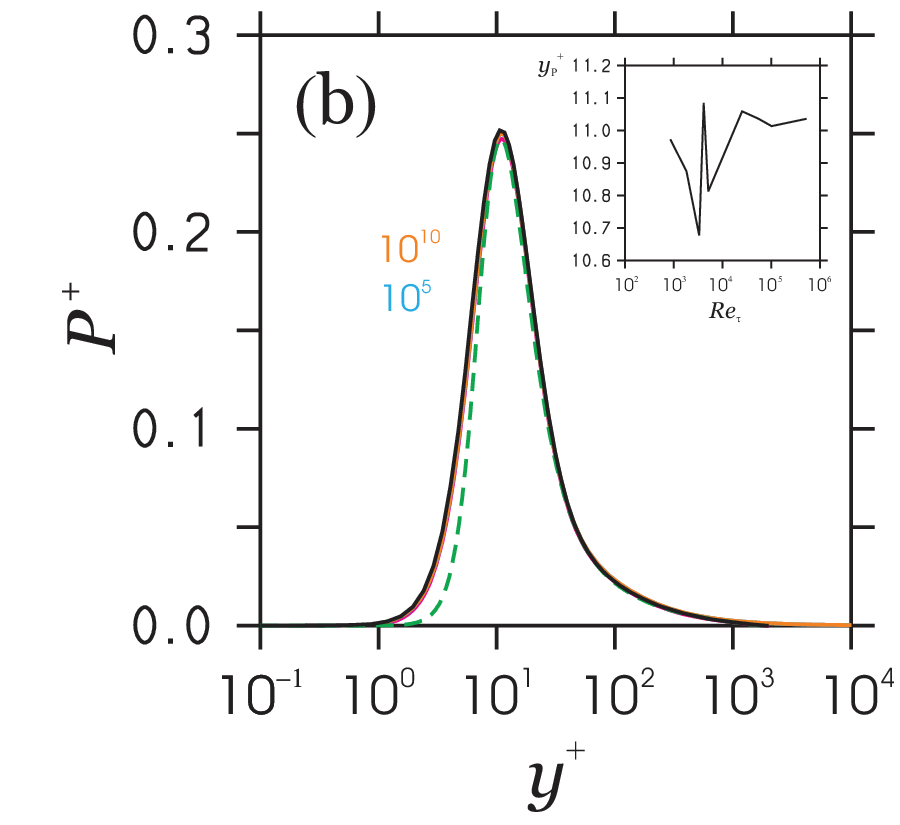} 
\includegraphics[width=\g1\textwidth]{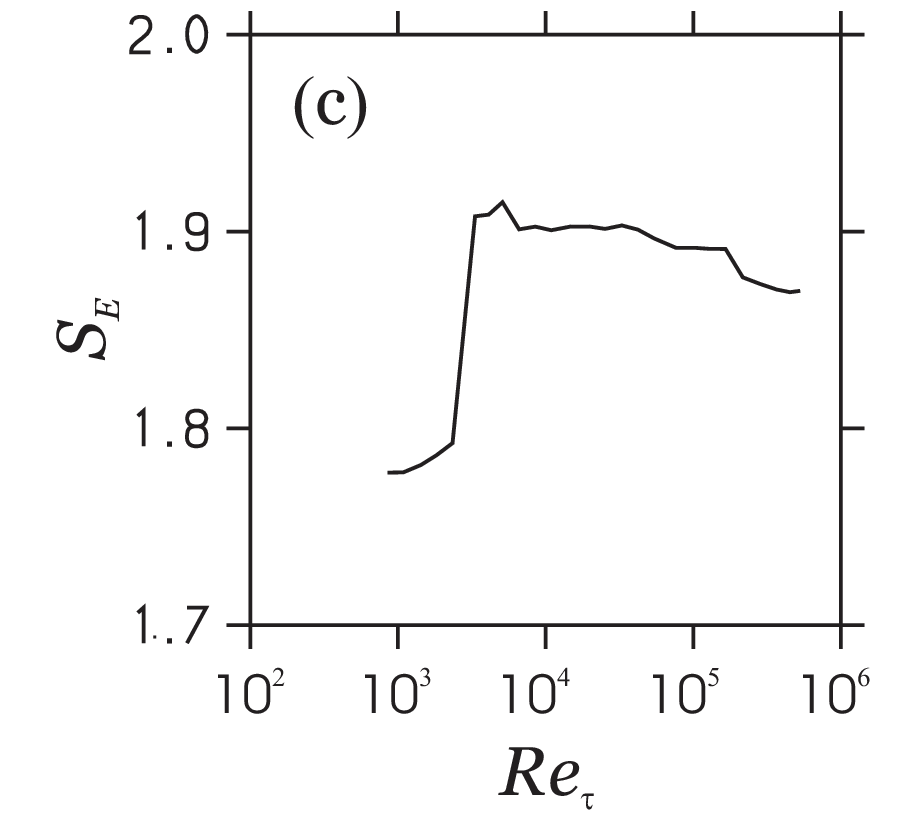} 
\linespread{1.}    \vspace{-0.3cm}
\caption{
Normalized turbulent viscosity $\nu_{t}^{+}=-\langle u'v'\rangle^{+}/S^{+}$ (a) and turbulence production $P^{+}=-\langle u'v'\rangle^{+}S^{+}$ (b) predictions: pipe flow DNS data~\citep{DNS-pipe} (black lines), Cantwell model results~\citep{Cant-19} (dashed green lines, $Re_{\tau}=1825$), and PVM results. PVM predictions for $Re_{\tau}=(2003, 10^{5}, 10^{10})$ are shown by magenta, cyan, and orange lines. The inset in (b) shows production peak positions according to Cantwell's model for the given $Re_{\tau}$.
Figure (c) shows the entropy $S_E$ according to Cantwell's model for the given $Re_{\tau}$. 
 }                                                                               
\label{fig:Cant}
\vspace{0.cm}
\end{figure*}

A nonuniversal velocity model was presented recently by~\citet{Cant-19}, see also~\citet{Cant-22}. 
The model uses classical mixing length theory and an ad hoc  model for the mixing length $\lambda$,
\vspace{-0.4cm}
\begin{equation}
-\langle u'v'\rangle^{+}=\lambda^{2} S^{+2}, \qquad \lambda=\frac{ky^+[1-e^{-(y^+/a)^m}]}{[1+y^{n}/b^n ]^{1/n}}. 
\vspace{-0.5cm}
\label{eq:Cant-1} 
\end{equation}
By using Eq.~(\ref{eq:Cant-1}) in the momentum equation $S^{+}-\langle u'v'\rangle^{+}=M$ we can find a model for $S^{+}$,
\vspace{-0.4cm}
\begin{equation}
S^{+}=-\frac{1}{2\lambda^{2}}+\frac{1}{2\lambda^{2}} \Big[1+4\lambda^2(1-y) \Big]^{1/2}, 
\vspace{-0.3cm}
\label{eq:Cant} 
\end{equation}
where $M=1-y$ for the pipe flow considered. This model involves five adjustable parameters, $k, a, m, b, n$, where $k$ corresponds to the von K\'{a}rm\'{a}n constant. Fig. 15d in Ref.~\citep{Cant-19} reveals that there is no strict log-law region with the influence of wall and outer length scales on the intermediate region of the velocity profile persisting to all $Re_{\tau}$~\citep{Cant-19}. A comparison of model predictions of the normalized turbulent viscosity $\nu_{t}^{+}=-\langle u'v'\rangle^{+}/S^{+}$ and turbulence production $P^{+}=-\langle u'v'\rangle^{+}S^{+}$ with DNS data and PVM predictions is shown in Fig.~\ref{fig:Cant}: it may be seen that the model does not accurately reflect the flow structure. An interesting model feature is the following. The $y$ contribution in $S^{+}$ is very small for sufficiently high $Re_{\tau}$, its influence decreases with increasing $Re_{\tau}$. For $Re_{\tau}>5000$, there is no observable difference anymore between  $S^{+}$ calculated by Eq.~(\ref{eq:Cant}) and $S^{+}$ calculated by neglecting $y$. 
In this case, analysis of $dP^{+}/d\lambda=0$ shows that the $P^{+}$ maximum appears at $\lambda=2^{1/2}$ (at $S^+=1/2$), which provides a maximum $P^{+}=1/4$. A similar PVM analysis leads to a maximum at $\ell=2^{1/2}$ corresponding to $S_{1}^{+}+S_{2}^{+}=1/2$ and also $P^{+}=1/4$.

The nonuniversality of Cantwell's model~\citep{Cant-19} is reflected by the need to provide $k, a, m, b, n$ (which are determined from an analysis based on the whole velocity profile) as functions of $Re_{\tau}$. 
The reason for this nonuniversality is the empirical introduction of $Re_{\tau}$ effects via $y=y^+/Re_{\tau}$ in $\lambda$ leading to an unphysical dependence of $\langle u'v'\rangle^{+}$ on $Re_{\tau}$. 
The following provides evidence for this claim. 
\vspace{-0.1cm}
\begin{compactenum} [O1.]
\item  
The simplest way to support this claim is the comparison of Eq.~(\ref{eq:Cant-1}) with PVM consequences: Eq.~(\ref{eq:Cant-1}) does not ensure a self-similar structure of the Reynolds shear stress in contrast to corresponding PVM implications~\citep{Fluids24-Law}. 
\item  
The turbulence production peak position is known to be $y^+=11.07$, unaffected by $Re_{\tau}$ for sufficiently high $Re_{\tau}$~\citep{JOT-18a, JOT-18b}: see Fig.~\ref{fig:Cant}. For $Re_{\tau}>5000$, Cantwell's model provides the production peak position at $\lambda=2^{1/2}$. According to the definition $\lambda=ky^+[1-e^{-(y^+/a)^m}]/([1+y^{n}/b^n ]^{1/n})$
 and the $Re_{\tau}$ dependence of model coefficients, this implies $y^+$ peak positions which vary (randomly) with $Re_{\tau}$: see the inset in Fig.~\ref{fig:Cant}b. This behavior is unphysical and in contrast to DNS and experimental results. 
\item  More specifically, the model's entropy is given by $S_E=1-ln(\kappa)=1-ln(k)$~\citep{JOT-18a}. We find, therefore, random entropy changes for each $Re_{\tau}$  and flow (see Fig.~\ref{fig:Cant}c). This is unphysical, the entropy needs to be the same under physically equivalent conditions.
\end{compactenum}

\begin{figure*}[t] 
\center
\includegraphics[width=\g1\textwidth]{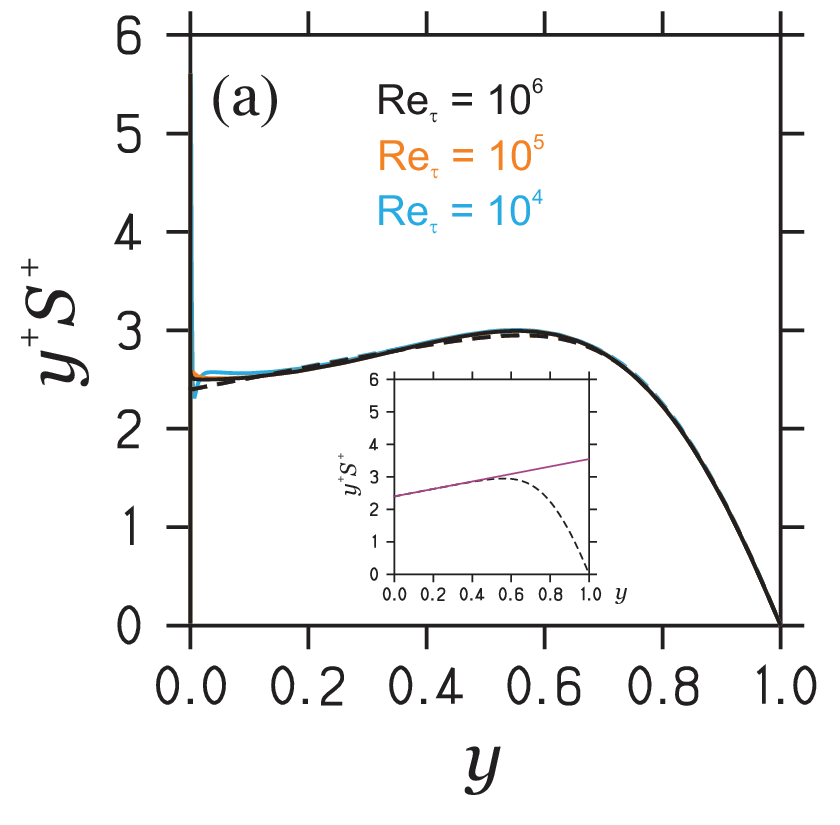} 
\includegraphics[width=\g1\textwidth]{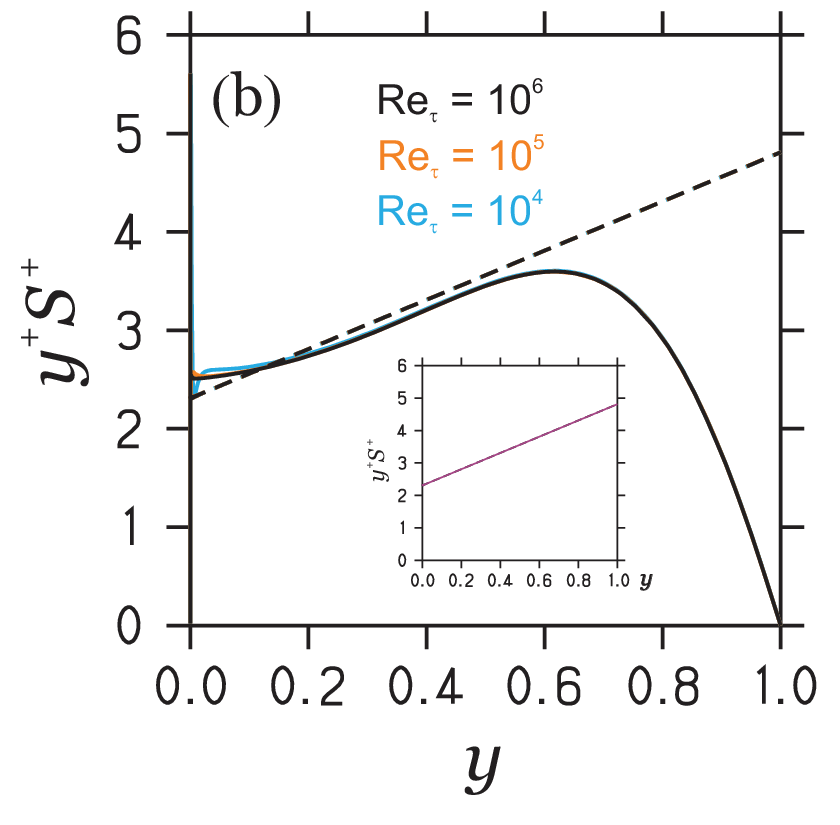} 
\includegraphics[width=\g1\textwidth]{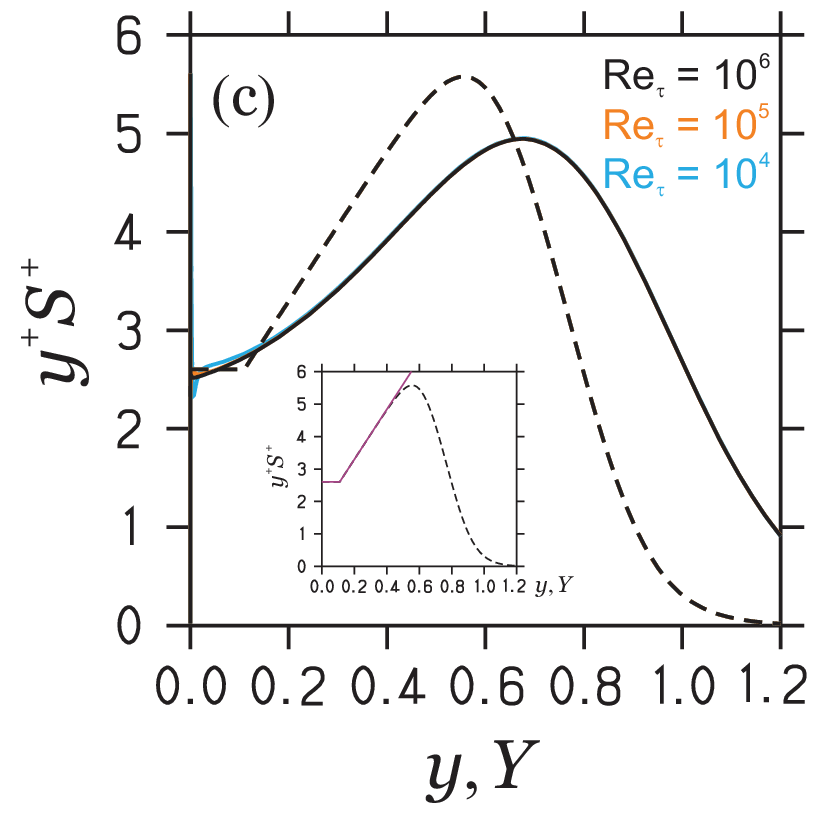} 
\linespread{1.}    \vspace{-0.3cm}
\caption{
The log-law indicator $y^{+} S^{+}$ in outer scaling obtained by the PVM (solid lines) versus MN assumptions~\citep{Monke-23} (dashed lines) at the given $Re_{\tau}$ for (a) channel flow, (b) pipe flow, and (c) TBL. For the TBL, the MN assumption is shown depending on the boundary layer thickness $Y$. $Re_{\tau}$ effects are hardly visible. The insets show the corresponding MN assumptions~\citep{Monke-23}:  $y^{+} S^{+}$ (dashed lines) are shown for $Re_{\tau}=10^{6}$ (there is no $Re_{\tau}$ effect). 
Also shown are corresponding linear profiles $1/0.417+1.15y$, $1/0.433	+2.5y$, $1/0.384$ and $1/0.384+7.7(Y-0.11)$ (purple lines), respectively. 
}                                                                               
\label{fig:Monk2}
\vspace{0.cm}
\end{figure*}

Another nonuniversal velocity model was presented recently by Monkewitz and Nagib (MN)~\citep{Monke-23}. Figure~\ref{fig:Monk2} demonstrates the model concept by a comparison with PVM results: MN approximate the log-law indicator $y^{+} S^{+}$ in regard to outer scaling. By following MN, the TBL results are shown in dependence on the boundary layer thickness $Y$, which differs from the 99\% boundary layer 
thickness~\citep{Monke-23}. 
The structure of MN assumptions is illustrated in the insets of Fig.~\ref{fig:Monk2}: linear functions are used to characterize $y^{+} S^{+}$.

\begin{figure*}[t] 
\center
\includegraphics[width=\g1\textwidth]{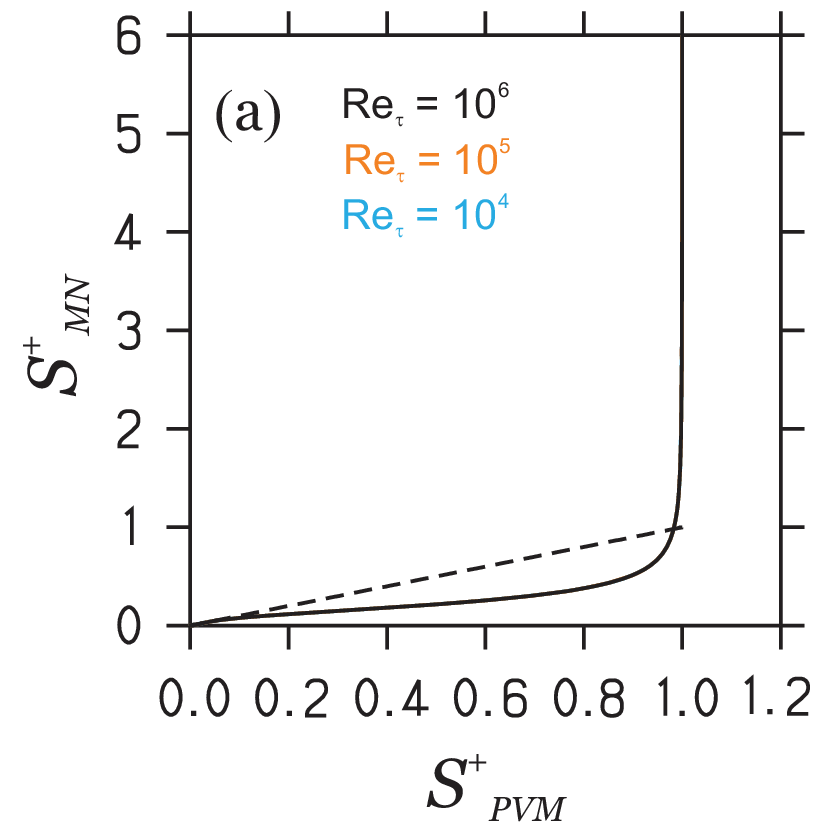} 
\includegraphics[width=\g1\textwidth]{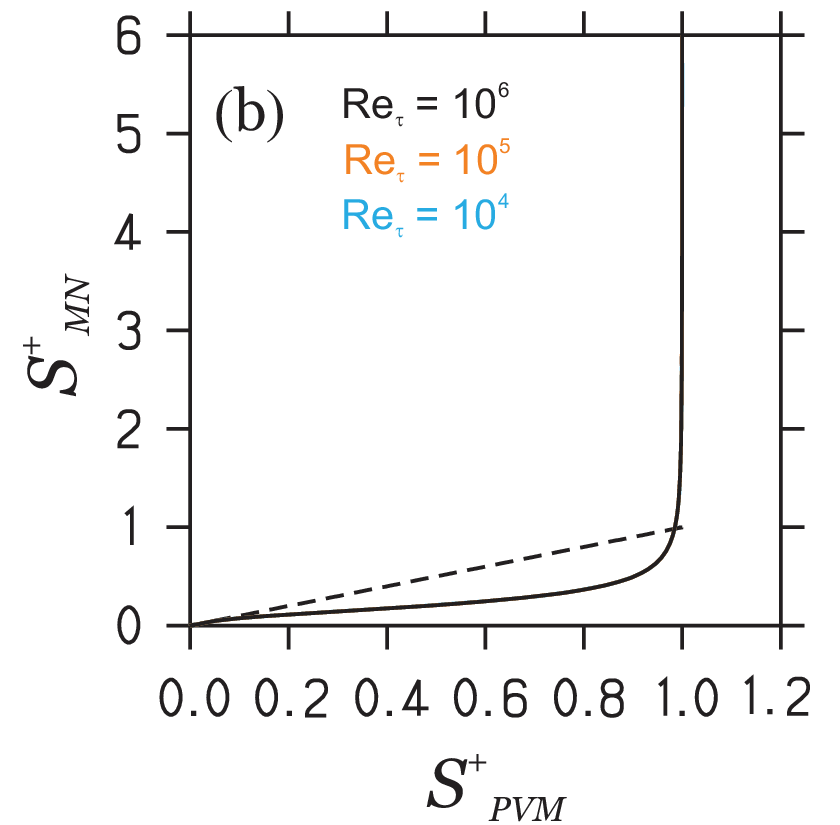} 
\includegraphics[width=\g1\textwidth]{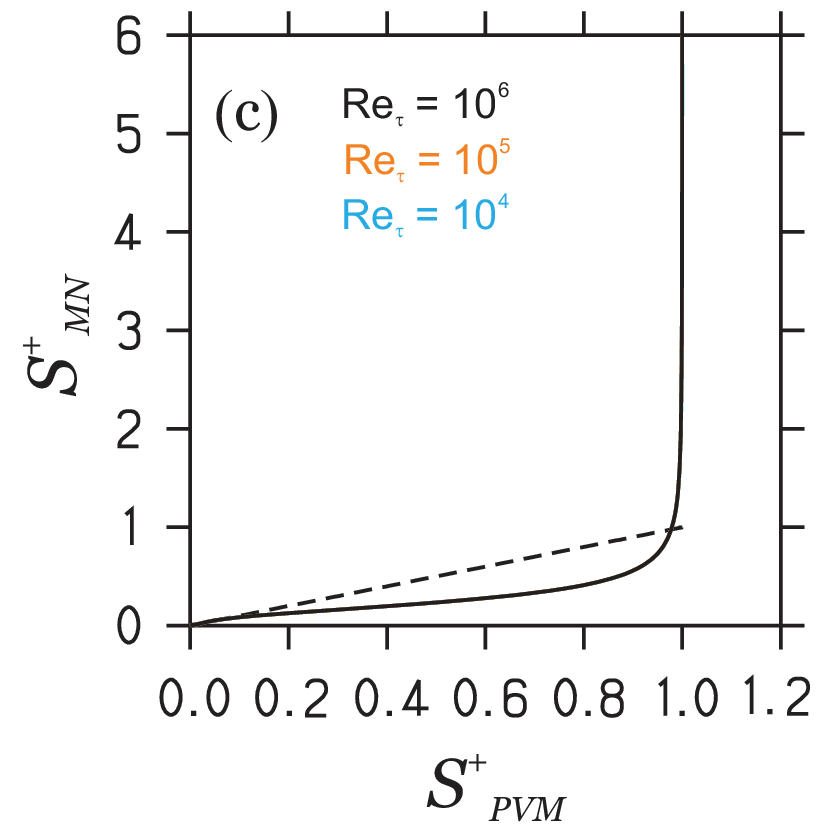} 
\linespread{1.}    \vspace{-0.3cm}
\caption{
The correlation of $S^{+}$ obtained by the PVM ($S_{PVM}^{+}$) and MN ($S_{MN}^{+}$) is shown for the given $Re_{\tau}$ for (a) channel flow, (b) pipe flow, and (c) TBL. There is no visible $Re_{\tau}$ effect. The dashed lines show the expected 1:1 relationships.
}                                                                               
\label{fig:Monk3}
\vspace{-0.cm}
\end{figure*}

In regard to MN's model, there seems to be a reasonable agreement between the PVM and MN assumptions. 
However, a closer look reveals an unphysical model behavior. 
\begin{compactenum}[O1.] 
\setcounter{enumi}{3} 
\item  
MN use flow-dependent outer scaling $y^{+} S^{+}$ variations (which scale with $y$) to determine $\kappa$ based on $y^{+} S^{+}\rightarrow 1 /\kappa$ for $y\rightarrow 0$. 
But the PVM shows that the contribution of outer scaling variations given by $y^{+} S_3^{+}$ becomes negligible compared to $1 /\kappa$ for $y\rightarrow 0$ (see Fig.~\ref{fig:structure}), i.e. the value of $\kappa$  is independent of $y^{+} S_3^{+}$ contributions. 
Hence, $\kappa$ cannot be determined by the analysis of outer scaling $y^{+} S^{+}$ variations: 
$\kappa$  characterizes inner scaling variations.

\item  MN present models for $S^{+}$ and $U^{+}=\int_0^{y^{+}}S^{+}(s)ds$. The MN assumptions imply that both $S^{+}$ and  $U^{+}$ diverge for $y\rightarrow 0$ (which is the regime used by MN to determine $\kappa$). There is no way to determine $\kappa$ if the underlying $S^{+}$ and  $U^{+}$ do not exist for $y\rightarrow 0$. 
\item  Fig.~\ref{fig:Monk3} shows the correlation of $S^{+}$ obtained by the PVM ($S_{PVM}^{+}$) and MN ($S_{MN}^{+}$). Despite remarkable discrepancies, the most relevant observation is that $S_{MN}^{+}$ can exceed unity. 
Combined with the momentum equation $-\langle u'v'\rangle^{+}=M-S^{+}$, we see that the MN model allows values of $-\langle u'v'\rangle^{+}$ outside $0 \leq -\langle u'v'\rangle^{+}\leq 1$, i.e. the MN model violates stress realizability requirements~\citep{JPAS-20}. 
\end{compactenum}


\begin{figure*}[t] 
\center
\includegraphics[width=\g1\textwidth]{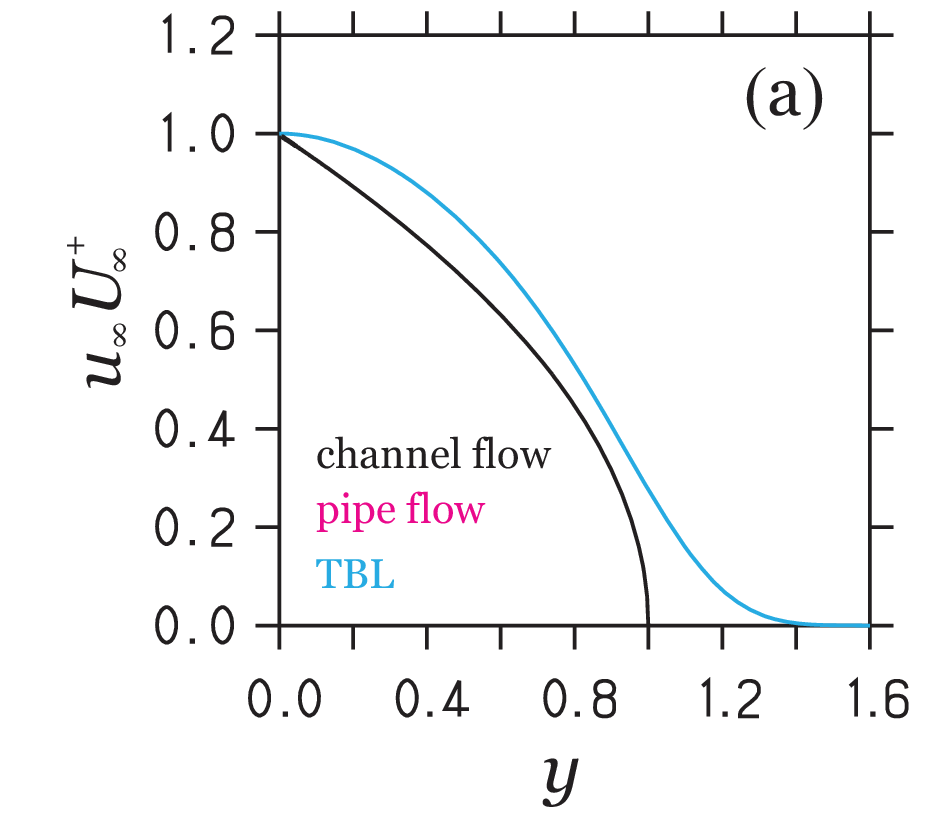} 
\includegraphics[width=\g1\textwidth]{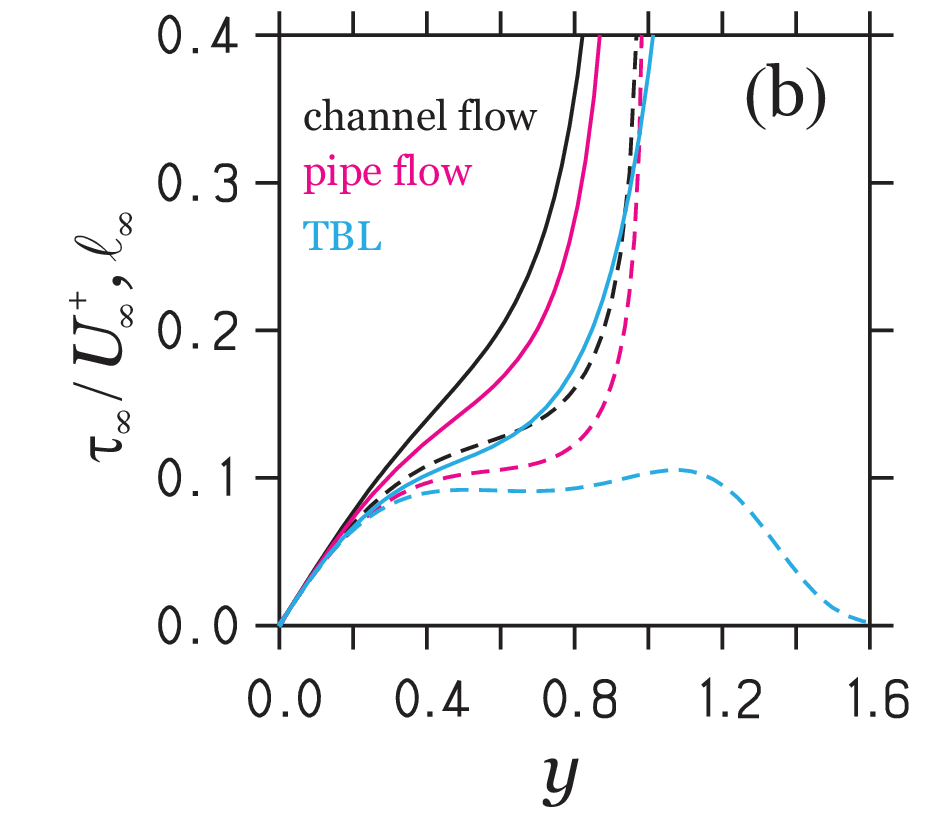} 
\includegraphics[width=\g1\textwidth]{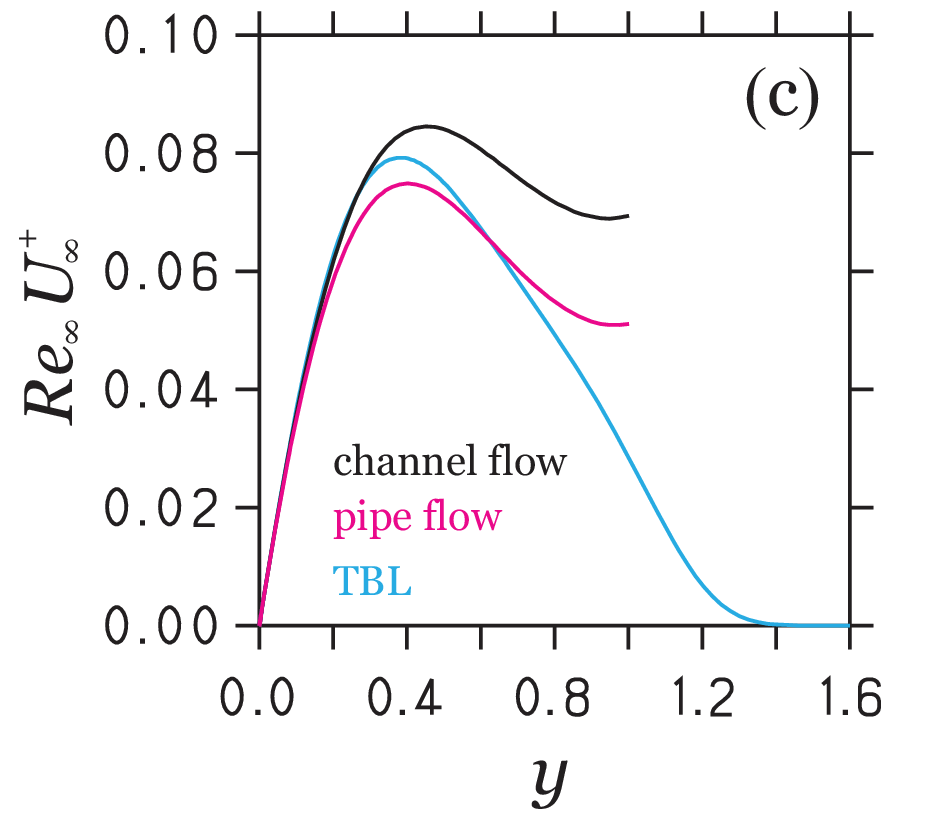} 
\linespread{1.}    \vspace{-0.4cm}
\caption{
Asymptotic outer scaling for the three flows considered: (a) turbulence velocity scale $u_*=u_\infty U_{\infty}^+$, (b) turbulence time $\tau_*/Re_{\tau}=\tau_\infty /U_{\infty}^+$ and length scales $\ell_*/Re_{\tau}=\ell_\infty $ (dashed lines), and (c)  $Re_{*}/Re_{\tau}=Re_{\infty} U_{\infty}^+$ . 
}                                                                               
\label{fig:AS-us}
\vspace{0.cm}
\end{figure*}

\begin{table*}[t] 
\footnotesize
\centering
\scalebox{1.0}{
\begin{tabular}{l c c c c c c c c c c c c c c c c c}  
\hline\hline  \vspace{0.05cm}
\rule{0pt}{2.5ex} \hspace{-0.25cm} & Outer-scale variables  & Inner-scale variables  & Outer-scale profiles\\[.1ex]  \hline  

\rule{0pt}{2.5ex} \hspace{-0.25cm} scaling velocity \& length & $U_{\infty}, \delta$  & $u_{\tau}, \delta$ & \\[.ex]

\rule{0pt}{2.5ex} \hspace{-0.25cm}  Reynolds number & $Re=U_{\infty}\delta/\nu=U_{\infty}^+ Re_{\tau}$  & $Re_{\tau}=u_{\tau}\delta/\nu$ &  \\[.1ex]

\rule{0pt}{2.5ex} \hspace{-0.25cm}  turbulence velocity scale & $u_{\infty}=\sqrt{-\langle u'v'\rangle^{+}}/U_{\infty}^+=u_{*}/U_{\infty}^+$  & $u_{*}=\sqrt{-\langle u'v'\rangle^{+}}$ &  $u_{\infty}= M^{1/2}/U_{\infty}^+$ \\[.ex]

\rule{0pt}{2.5ex} \hspace{-0.25cm} turbulence  time scale & $\tau_{\infty}=U_{\infty}^+/(S^+Re_{\tau})=\tau_{*}U_{\infty}^+/Re_{\tau}$  & $\tau_{*}=1/S^+$ & $\tau_{\infty}=\kappa y U_{\infty}^+/[1+\kappa y^+ S_3^+]$ 
  \\[.ex]

\rule{0pt}{2.5ex} \hspace{-0.25cm} turbulence length scale & $\ell_{\infty}=u_{\infty}\tau_{\infty}=\ell_*/Re_{\tau}$  & $\ell_{*}=u_{*}\tau_{*}=\sqrt{-\langle u'v'\rangle^{+}}/S^+$ & $\ell_{\infty}=\kappa y M^{1/2}/[1+\kappa y^+ S_3^+]$ \\[.ex]

\rule{0pt}{2.5ex} \hspace{-0.25cm}  turbulence $Re$ & $Re_{\infty}=u_{\infty}\ell_{\infty}=Re_{*}/(U_{\infty}^+ Re_{\tau})$  & $Re_{*}=u_{*}\ell_{*}=-\langle u'v'\rangle^{+}/S^+$ & $Re_{\infty}=\kappa y M/[1+\kappa y^+ S_3^+] 
/U_{\infty}^+$ \\[.5ex]
\hline\hline  
\end{tabular}
}
\vspace{-0.2cm}\centering \caption{Overview of inner and outer scaling variables, and outer scale profiles. Here, 
$U_{\infty}^{+}=5.03+\kappa^{-1}ln(Re_{\tau}/K)$ is the centerline/ freestream maximum velocity. $Re_{*}=\nu_t^+$ is equivalent to the inner-scale turbulence viscosity.}
\label{tab:scalings}
\end{table*}

\begin{figure}[t] 
\center
\includegraphics[width=0.22\textwidth]{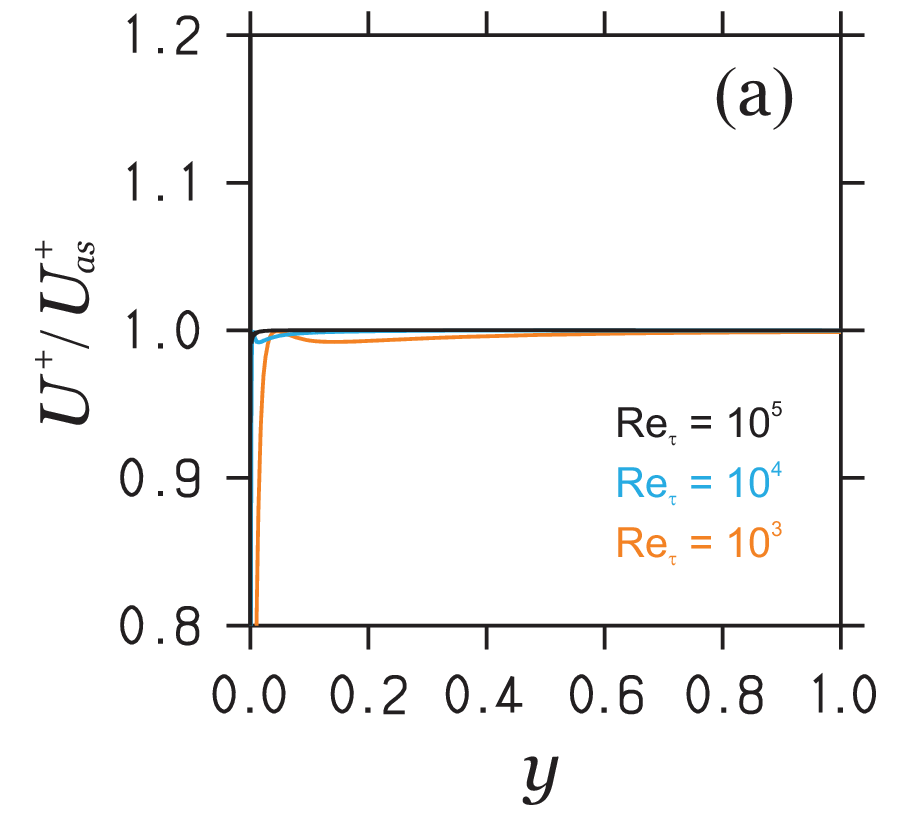} 
\includegraphics[width=0.22\textwidth]{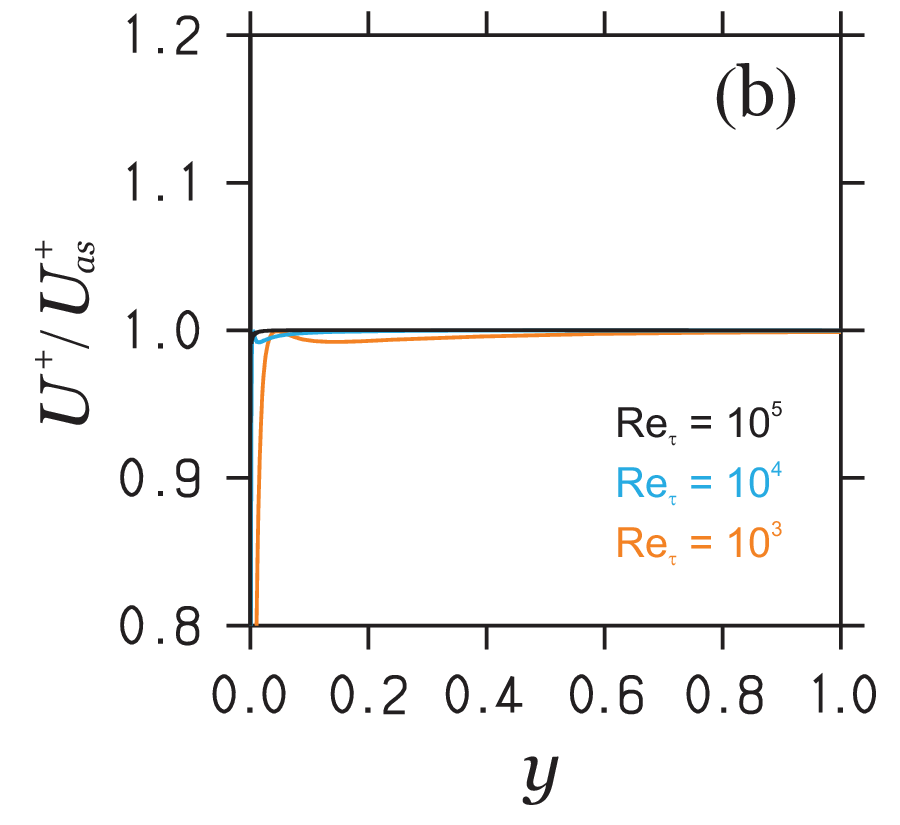} 
\linespread{1.}    \vspace{-0.4cm}
\caption{
Asymptotic outer velocity scaling of $U^+/U^+_{as}$ with $Re_{\tau}$ along $y$: (a) channel flow, (b) pipe flow.
}                                                                               
\label{fig:AS-VPlus}
\vspace{-0.4cm}
\end{figure}

\section*{ASYMPTOTIC FLOW STRUCTURE}

To illustrate the advantage of the universal PVM we consider implications in regard to the asymptotic structure of canonical wall-bounded turbulent flows. The asymptotic variation of $U^+$ implied by the PVM is given by 
\vspace{-0.4cm}
\begin{equation}
U^+_{as}=U_{\infty}^+ +\frac{1}{\kappa}ln\Big( \frac{Ky}{Ky+w} \Big)
=\frac{1}{\kappa}  ln(y^+)+5.03 +U_3^+,
 \vspace{-0.3cm}
\label{eq:scaling-y-asymp} 
\end{equation}
where $U_{\infty}^+=5.03+\kappa^{-1}ln(Re_{\tau}/K)$ and $U_3^+=-\kappa^{-1}ln(Ky+w)$ are used 
($U_3^+$  is the wake contribution to $U^+$ based on $S_3^+$).
As shown in Fig.~\ref{fig:AS-VPlus}, $U^+$ converges to this asymptotic scaling: at $Re_{\tau}=10^5$, there is hardly any visible difference between $U^+ /U^+_{as}$ and unity anymore. 
There is also hardly any difference between the flows considered (the TBL plot shows the same).
We note that the neglect of boundary effects (the neglect of $U_3^+$) implies the universal velocity log-law.

Characteristic properties of turbulence can be well studied by considering characteristic outer-scale velocity, time, and length scales $u_{\infty}$, $\tau_{\infty}$, and $\ell_{\infty}$ (which are only functions of $y$) as defined in Table~\ref{tab:scalings}. 
Figure~\ref{fig:AS-us} presents the corresponding asymptotic distributions of turbulence velocity scales, time scales, length scales, and turbulence Reynolds numbers for the three flows considered. Independent of specific distributions, the most relevant observation is that the turbulence asymptotically decays, as may be seen from the $Re_{\infty}$ trends under consideration of the fact that $Re_{\infty} \sim 1/U_{\infty}^+ \rightarrow 0$, where $U_{\infty}^{+}=5.03+\kappa^{-1}ln(Re_{\tau}/K)$. 
In correspondence to that we find that the turbulence velocity scale vanishes, $u_{\infty}\sim 1/U_{\infty}^+ \rightarrow 0$, and the time scale $\tau_{\infty}\sim U_{\infty}^+ \rightarrow \infty$. 
The structure of $Re_{\infty} U_{\infty}^+=Re_{*}/Re_{\tau}$ corresponds to expectations: for channel and pipe flow we see a damping-function type distributions along $y$ that approaches a constant Reynolds number at the centerline. For the TBL, the flow becomes laminar  under freestream conditions. 

The distribution of the length scale $\ell_{\infty}$ seen in Fig.~\ref{fig:AS-us}  is of particular interest. In contrast to the other variables ($Re_{\infty}$, $u_{\infty}$, $\tau_{\infty}$), 
$\ell_{\infty}$  is finite over most of the domain. In particular near the wall $\ell_{\infty}$ follows 
$\ell_{\infty}=\kappa y$ for all the flows considered. The latter provides strong support for the suitability of Prandtl's debated mixing length concept~\citep{Prandtl-25}. For channel and pipe flow $\ell_{\infty}$  diverges for $y\rightarrow 1$, the size of turbulence structures can become unbounded. For the TBL case we see that $\ell_{\infty}$  approaches zero under freestream conditions, which is consistent with the $Re_{\infty}$ behavior showing a flow laminarization.

\section*{SUMMARY}

The paper addressed the ongoing controversial debate about the universality or nonuniversality of the law of the wall. 
According to the observations O1-O6, the conclusion is that observed nonuniversality~\citep{Cant-19, Cant-22,  Monke-23}  is a consequence of model assumptions that are in conflict with physics, whereas a universal law of the wall implied by the PVM is found if physics requirements are honored.

Such differences have simple practical consequences. In contrast to nonuniversal models, the universal PVM model presented here enables (i) the reliable validation of computational methods under variable high $Re$ conditions where other methods are inapplicable, (ii) the design of more accurate turbulence models including $Re$ effects~\citep{Plaut-22}, and (iii) to establish a bridge between finite $Re$ observations and asymptotic structural theories of turbulence~\citep{Fluids24-As}. As an example for the latter we note the agreement between the universal von K\'{a}rm\'{a}n constant $\kappa=0.4$ presented here and $\kappa=1/(2\pi)^{1/2}=0.399$ derived by~\citet{Baumert-13} based on turbulence structure assumptions at infinite $Re$.

\section*{ACKNOWLEDGMENTS}

I would like to acknowledge support from the National Science Foundation (AGS, Grant No.~2137351, with Dr. N. Anderson as Technical Officer) and support from the Hanse-Wissenschaftskolleg.

\bibliographystyle{tsfp}


\end{document}